\documentclass[aip,graphicx,reprint]{revtex4-1}
\usepackage[cmex10]{amsmath}
\usepackage{ amssymb }
\usepackage[pdftex]{graphicx}
\usepackage{float}
\usepackage{multirow}
\usepackage{epstopdf}
\usepackage{color}

\draft 

\begin{document}

\title{Strain effect in highly-doped n-type 3C-SiC-on-glass substrate for mechanical sensors and mobility enhancement} 

\author{Hoang-Phuong Phan}
\email{hoangphuong.phan@griffithuni.edu.au}
\affiliation{Queensland Micro- and Nanotechnology Centre, Griffith University, QlD, Australia} 
\affiliation{Aeronautics and Astronautics Department, Stanford University, CA, USA}
\author{Tuan-Khoa Nguyen}
\affiliation{Queensland Micro- and Nanotechnology Centre, Griffith University, QlD, Australia}
\author{Toan Dinh}
\affiliation{Queensland Micro- and Nanotechnology Centre, Griffith University, QlD, Australia}
\author{Han-Hao Cheng}
\affiliation{Centre for Microscopy and Microanalysis, University of Queensland, QlD, Australia} 
\author{Fengwen Mu}
\affiliation{Department of Precision Engineering, The University of Tokyo, Tokyo, Japan} 
\author{Alan Iacopi}
\affiliation{Queensland Micro- and Nanotechnology Centre, Griffith University, QlD, Australia}
\author{Leonie Hold}
\affiliation{Queensland Micro- and Nanotechnology Centre, Griffith University, QlD, Australia}
\author{Tadatomo Suga}
\affiliation{Department of Precision Engineering, The University of Tokyo, Tokyo, Japan} 
\author{Dzung Viet Dao}
\affiliation{Queensland Micro- and Nanotechnology Centre, Griffith University, QlD, Australia}
\affiliation{School of Engineering, Griffith University, Qld, Australia}
\author{Debbie G. Senesky}
\affiliation{Aeronautics and Astronautics Department, Stanford University, CA, USA} 
\affiliation{Department of Electrical Engineering, Stanford University, CA, USA.}
\author{Nam-Trung Nguyen}
\affiliation{Queensland Micro- and Nanotechnology Centre, Griffith University, QlD, Australia}

\date{\today}

\begin{abstract}
This work reports the strain effect on the electrical properties of highly doped n-type single crystalline cubic silicon carbide (3C-SiC) transferred onto a 6-inch glass substrate employing an anodic bonding technique. The experimental data shows high gauge factors of -8.6 in longitudinal direction and 10.5 in transverse direction along the [100] orientation. The piezoresistive effect in the highly doped 3C-SiC film also exhibits an excellent linearity and consistent reproducibility after several bending cycles. The experimental result was in good agreement with the theoretical analysis based on the phenomenon of electron transfer between many valleys in the conduction band of n-type 3C-SiC. Our finding for the large gauge factor in n-type 3C-SiC coupled with the elimination of the current leak to the insulated substrate could pave the way for the development of single crystal SiC-on-glass based MEMS applications.
\end{abstract}


\maketitle 

The demand for electronics and sensors which can operate in extreme conditions has propelled the research into wide band gap materials such as III-nitride (e.g. GaN, AlN), silicon carbide (SiC) and diamond-like carbon (DLC) \cite{Senesky1,SiC_review1,SiC_review2,DLC}. Among these materials, SiC has emerged as an excellent semiconductor for Micro Electro Mechanical Systems (MEMS) applications owing to its large energy gap, varying from 2.3 eV to 3.4 eV, excellent chemical inertness, superior mechanical properties, and outstanding radiation tolerance \cite{SiC_Phan,SiC_Toan}. Numerous SiC technologies have been transitioned into commercial products, including bipolar transistors, Schottky diodes, field effect transistors, ultraviolet photodetectors and gas sensors \cite{SiC_app1,SiC_app2,SiC_app3,SiC_app4}. 

Although great progress has been made to commercialize SiC materials and devices, the material cost is still relatively expensive in comparison to the Si counterpart. One promising approach to bring down the cost of SiC materials is the development of epitaxial SiC nanothin films on a Si substrate \cite{3C-SiC_adv1,3C-SiC_adv2}. This approach can utilize the low cost and worldwide availability of Si substrate, as well as take advantage of Si-orientated MEMS fabrication technologies. To date, high quality and large scale SiC wafers grown on Si have been reported \cite{3C-SiC_app1}. Applications of SiC-on-Si have also been demonstrated such as mechanical sensors, chemical sensors, bio devices \cite{3C-SiC_app2,3C-SiC_app3,3C-SiC_app4}. Nevertheless, when subjected to high temperature, the current leakage from SiC to the Si substrate could adversely affect the performance of the SiC sensing element. Therefore, a number of efforts have been made to transfer SiC films from SiC/Si wafers onto electrically insulating substrates \cite{bonding1}. Most of the works reported in the literature are based on SiO$_2$ to SiO$_2$ direct bonding or  Si to poly Si diffusion bonding techniques to form an oxide layer sandwiched between SiC and Si \cite{bonding2,bonding3,bonding4}. However, these methods require an ultra-smooth buffer layer such as SiO$_2$, formed by sputtering/oxidation and polishing or SiO$_2$/poly-Si which typically complicates the process. Furthermore, the Si layer in the bonded wafers generally absorbs visible wavelengths, hindering the merit of high-optical-transmittance in bulk SiC materials. Employing the anodic bonding method, we have recently demonstrated large scale single crystal 3C-SiC films transferred onto a glass substrate \cite{bonding_Phan}. The bonded wafers possess robust adhesion strength, high visible-wavelength transmittance, and excellent electrical insulation to the substrate. In addition, the transferred SiC nanothin films have also been proven to be bio-compatible which enabled the direct culture of bio-cells. Furthermore, the large scale of 6-inch in the diameter also makes the single crystal 3C-SiC/glass wafers a promising platform to develop low-cost mass production devices for a wide range of MEMS applications. 

 This work elucidates the strain effect on the electrical conductance of highly doped n-type 3C-SiC nano films bonded onto glass through experimental characterizations and the first principle calculation. The experimental results show a negative gauge factor of -8.6, which is in solid agreement with theoretical analysis based on the many valleys electron transfer phenomenon. The insight obtained from this study will be the foundation for the development of SiC/glass mechanical sensing applications as well as opening up opportunities for further investigation into strain engineering in SiC.

\begin{figure*}[t!]
\centering
 \includegraphics[width=6.6in]{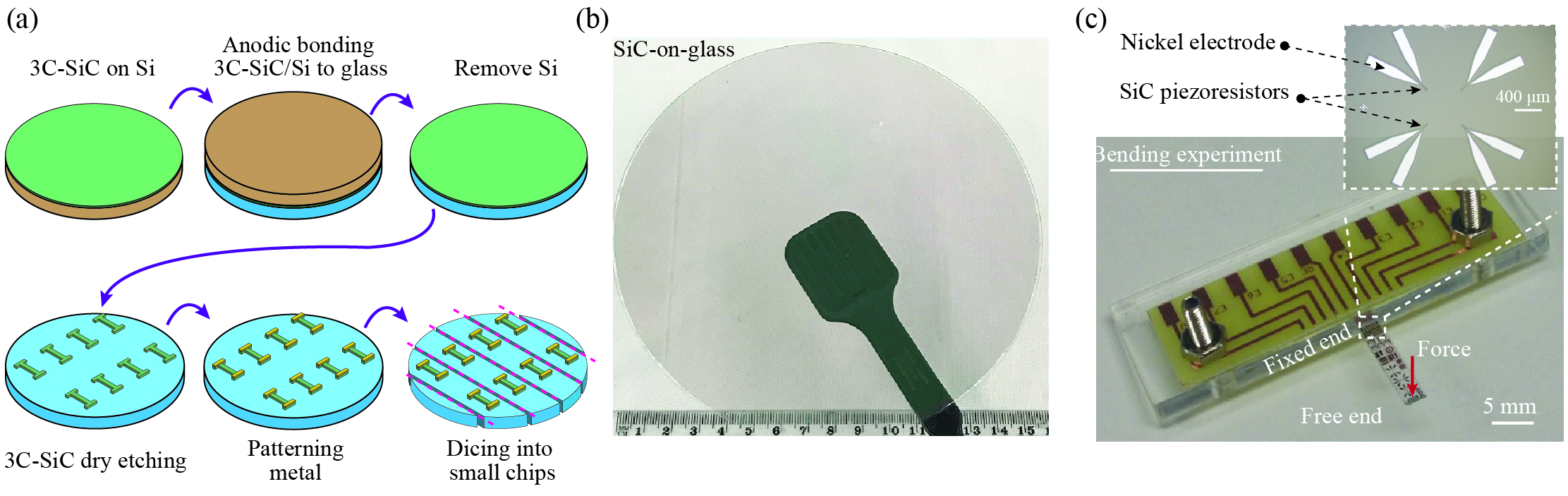}
 \caption{Fabrication of SiC/glass piezoresistors. (a) The concept of anodic bonding process and lithography on the SiC/glass platform; (b) Photograph of a full 6-inch SiC/glass wafer; (c) Photograph of a SiC/glass cantilever used in the bending experiment (In set: Microscope image of SiC piezoresistors aligned in longitudinal and transverse [100] orientations)}
\label{fig:fig1}
 \end{figure*}

To conduct this experiment, single-crystalline 3C-SiC thin films were grown on Si substrates using a low pressure chemical vapor deposition process (LPCVD) \cite{growth1,growth2}. Silane (SiH$_4$) and propene (C$_3$H$_6$) were employed as the precursors in the hot-wall furnace at 1250$^\circ$C. Additionally, ammonia (NH$_3$) was utilized as the \emph{in situ} dopant, forming highly-doped n-type 3C-SiC. The thickness of the initial 3C-SiC film was 700 nm, measured using a NANOMETRICS \texttrademark Nanospec/AFT 210. The film was further smoothed by etching back the SiC layer to a thickness of 600 nm to improve the contact surface between SiC and glass during wafer bonding. Prior to the bonding process, piranha cleaning and oxygen plasma was performed to remove contamination from the surface of glass and SiC/Si wafers. The smoothed 6-inch SiC/Si wafers were then subjected to the anodic bonding process, performed using an EVG\texttrademark 520IS hot embosser under a compressive force between 1.5 and 2.5 kN and bias voltage between -200 to -1000 V at a temperature of 400$^\circ$C. Subsequently, the top Si layer with a thickness of 650 $\mu$m was completely removed in KOH at 80$^\circ$C. Next, a two-mask fabrication process was carried out to fabricate SiC resistors using Inductive Coupled Plasma with an etching rate of 100nm/min. Finally, the SiC on glass samples was diced into cantilevers with dimensions of 0.51 mm$\times$3.5 mm$\times$20 mm for the subsequent bending experiment, as illustrated in Fig \ref{fig:fig1}(a). 

Figure \ref{fig:fig1}(b) shows a 6-inch 3C-SiC/glass wafer, exhibiting excellent transparency. The optical characterization and the bonding strength of the films can be found elsewhere \cite{bonding_Phan}. The large scale of the bonded wafers could significantly reduce the material cost as well as allow mass production of SiC/glass based devices by employing a batch fabrication process. 
 Figure \ref{fig:fig1}(c) shows a photograph of SiC/glass cantilever for the bending experiment, in which the U-shaped SiC piezoresistors with dimensions of 8 $\mu$m$\times$200 $\mu$m$\times$600 nm were patterned in the vicinity of the fixed end of the cantilever in order to obtain high tensile strain. In this work, we employed the SiC piezoresistors aligned along the longitudinal and transverse [100] directions to investigate the strain effect on electrical conductance, Fig. \ref{fig:fig1}(c) inset.
\begin{figure}[b!]
\centering
 \includegraphics[width=3.5in]{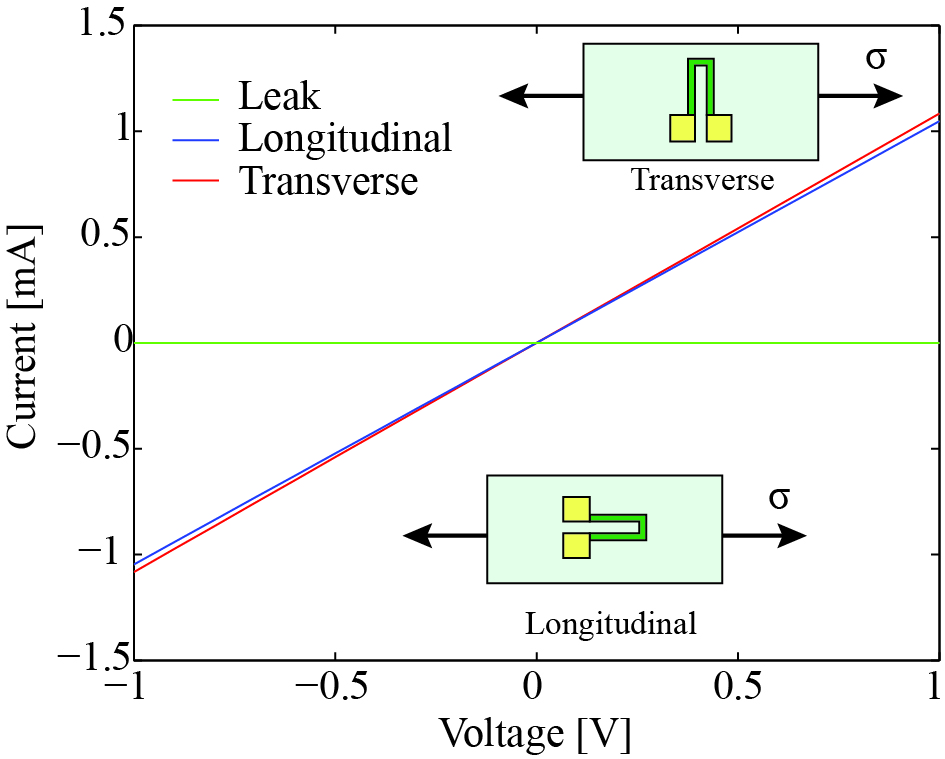}
 \caption{Current-Voltage (I-V) characteristic of the SiC piezoresistors. The current leakage was found to be in pA order, which is six orders of magnitude smaller than the current of the SiC resistors.}
\label{fig:fig2}
 \end{figure}

The carrier concentration of the bonded 3C-SiC films was found to be approximately 10$^{19}$ cm$^{-3}$, using a hot probe technique \cite{}. The 
current-voltage (I-V) characteristic of the SiC piezoresistors was measured using Agilent\texttrademark B1500A, indicating that a good Ohmic contact was formed between Ni and n-type 3C-SiC. In addition, the similarity in the resistances of longitudinal and transverse SiC piezoresistors also indicates the uniformity of the carrier concentration in the highly doped film. Furthermore, as the SiC was transferred onto glass, no current leakage was observed from the sensing layer to the substrate, as illustrated in Fig. \ref{fig:fig2}.  
\begin{figure*}[t!]
\centering
 \includegraphics[width=5.2in]{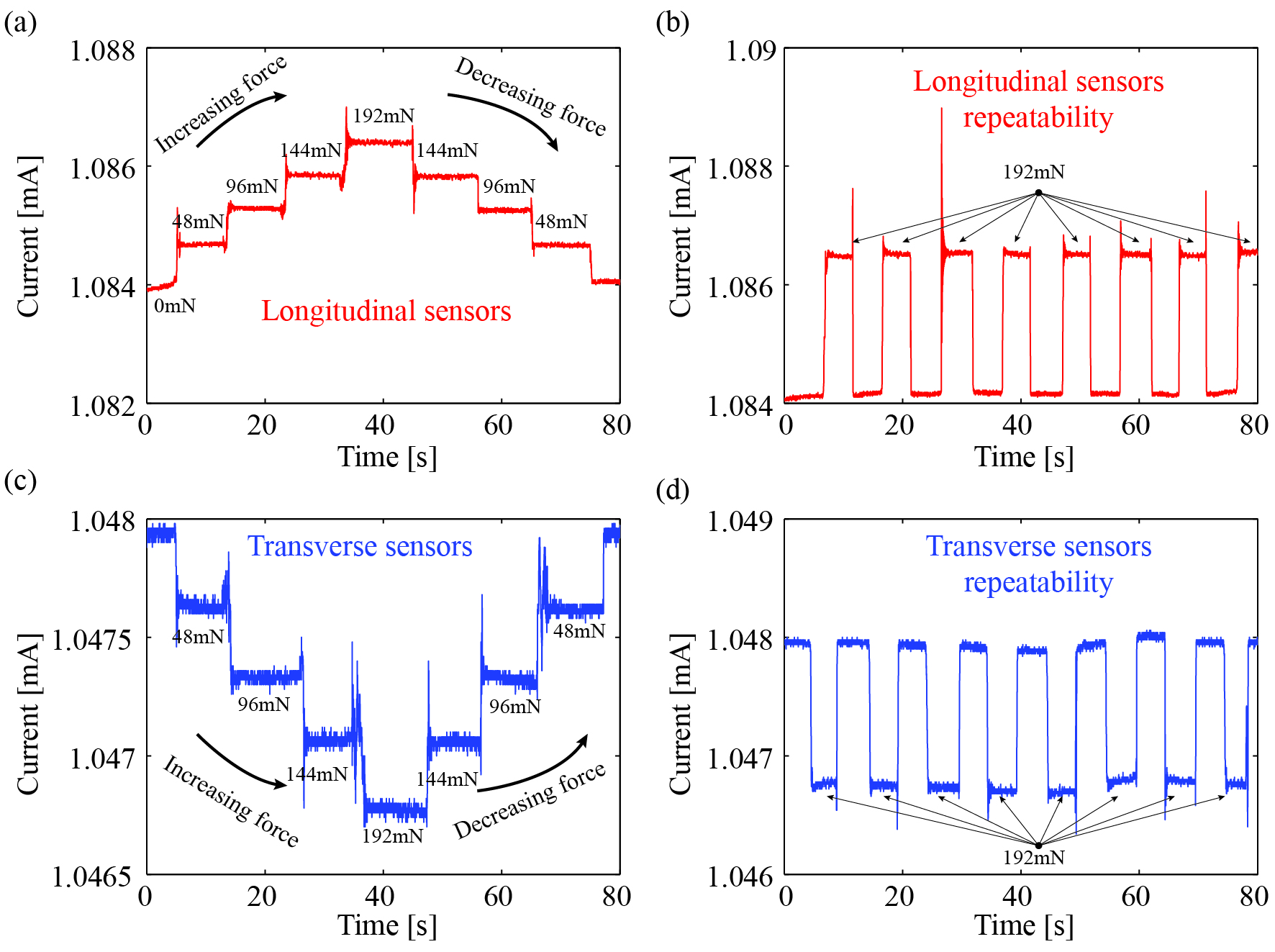}
 \caption{Characterization of the piezoresistive effect in n-type 3C-SiC on glass using the bending beam method. The change in the current of the longitudinal resistor (a) and (b); and of the transverse resistor (c) and (d) under strain and at a constant applied voltage of 1 V. The spikes were caused by the uncertainty of the experimental setup during loading/unloading forces.}
\label{fig:fig3}
 \end{figure*}

The bending beam method was applied to induce strain into the SiC piezoresistors, in which a SiC/glass cantilever was fixed, while the other free end was deflected downward by external forces. In addition, to avoid the effect of the boundary condition in the bending test, the clamped area was placed at least 2 mm far from the piezoresistors. 
The applied strain was estimated based on finite element analysis (FEA) using COMSOL Multiphysics\texttrademark). Accordingly, for the longitudinal resistance, nearly 100\% of the strain applied to the top surface of the glass cantilever was transmitted to the SiC layer. On the other hand, for the transverse resistor, due to the small ratio of width to length of approximately 1:11, only 48\% of applied strain was transmitted to the SiC layer (See supplementary online information). This phenomenon of small strains induced into transverse resistors has also been reported in the literature, which is more profound for the case of nanowires with extremely small width to length ratio \cite{strain_NWs}.  
Based on the FEA,  when varying the applied force from 0 to 192 mN, the strains induced into longitudinal and transverse SiC piezoresistors were estimated to vary from 0 to 310 ppm, and from 0 to 140 ppm, respectively.

The response of the n-type 3C-SiC piezoresistors under applied strain was then investigated by monitoring the change in the output current under a constant applied voltage of 1 V, as shown in Fig. \ref{fig:fig3}(a). Accordingly, when increasing the tensile strain, the current passing through the SiC piezoresistors also increased. This characteristic in n-type 3C-SiC is different from p-type 3C-SiC as well as metals where the current decreases under tensile strain \cite{SiC_Phan,SiC_app4}. The increase in the current in the n-type 3C-SiC also indicates the possibility to improve the mobility of n-type 3C-SiC on glass based devices by employing mechanical tensile strain. The change in the current also exhibits excellent repeatability after multiple bending cycles under an applied force of 192mN. Figures \ref{fig:fig3}(c)(d) plot the resistance change of the transverse resistor under applied loads, showing the opposite trend to that of the longitudinal resistor (i.e. resistance increased when increasing the tensile strain). Furthermore, it should be pointed out that under the same applied force to the SiC/Si cantilever, the change of the transverse resistance was relatively small in comparison to that of longitudinal resistance. This result is considerable due to the fact that a smaller strain was induced into the transverse resistor as described above. 

\begin{figure}[t!]
\centering
 \includegraphics[width=3.3in]{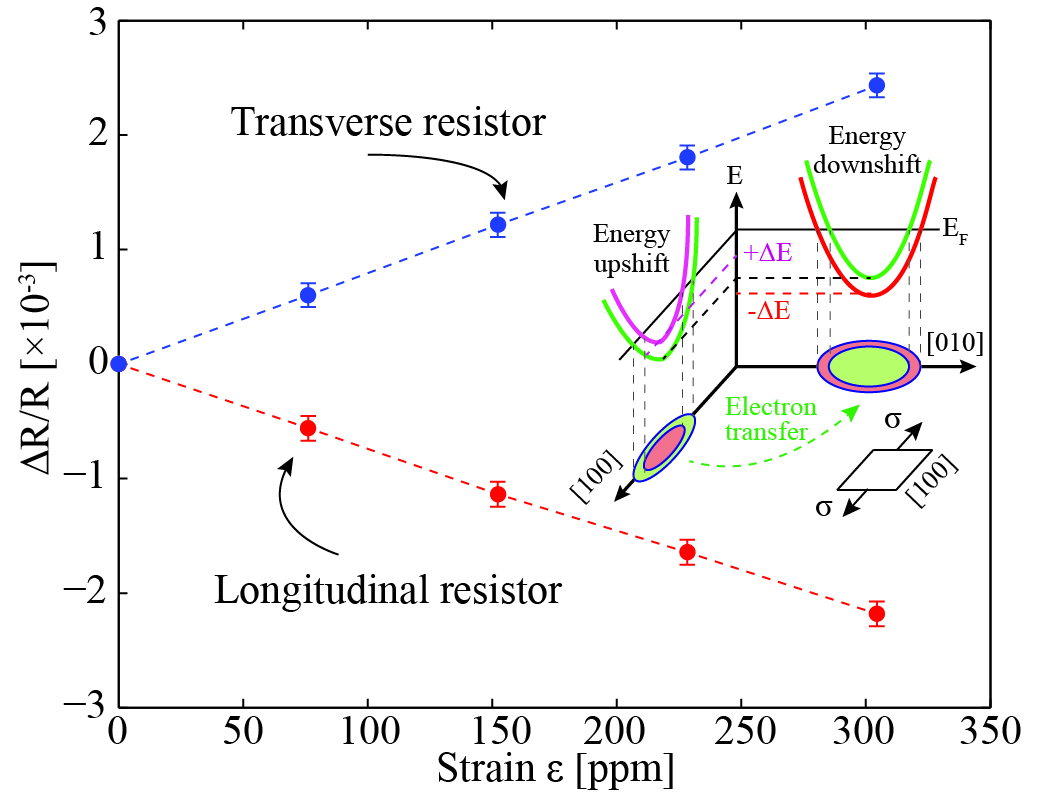}
 \caption{The relationship between resistance change in the n-type 3C-SiC resistances and applied strain. Inset: Schematic sketch of the electron transfer phenomenon.}
\label{fig:fig4}
\vspace{-2em}
 \end{figure}

Figure \ref{fig:fig4} shows the resistance change of the n-type 3C-SiC on glass against applied strain to obtain the gauge factor of the material ($GF=\frac{\Delta R/R}{\varepsilon}$).  A linear relationship between the resistance change and tensile strain was observed in both longitudinal and transverse resistors. This linearity is a favorable property for strain sensing applications. Based on the results from Fig. \ref{fig:fig4}, the longitudinal and transverse gauge factors of n-type 3C-SiC were found to be -8.6 and 10.5, respectively, which are at least 5 times larger than those of metals (See supplementary online information).

Employing the theory of strained n-type cubic semiconductors, we qualitatively explain the piezoresistance of n-type 3C-SiC on glass based on the electron transfer phenomenon \cite{Doll_book,Bir_book}. That is, under applied strain, the conduction bands of the n-type 3C-SiC are deformed, leading to the re-population of electrons, following the Boltzmann distribution. According to the first principle analysis of Nakamura \emph{et al.} on a n-type 3C-SiC nanosheet \cite{Nakamura1,Nakamura2}, under a tensile strain in [100] orientation, the energy valley in the [100] (i.e. the longitudinal valley) will shift up, while the energy valleys in [010] and [001] directions (i.e. the transverse valleys) will shift down. These shifts in the 6 energy valleys result in the re-population of free electrons, in which the electrons from the longitudinal direction (i.e. [100] orientation) will fill up the valley in transverse direction (i.e. [010] and [001] orientations). Moreover, since the transverse mobility of electron ($\mu_\perp$) is higher than that in the longitudinal direction ($\mu_\parallel$), more electrons in the [010] and [001] direction leads to a higher effective mobility. This hypothesis is consistent with our results that the tensile strain increases the conductivity of the n-type 3C-SiC, indicating the enhancement of electron mobility with the assumption that the total number of electrons remain constant under strain for highly doped semiconductors \cite{Sun_book}.

Next, we utilize the theoretical model of Kanda developed for Si--another cubic single crystal \cite{Kanda1} to numerically estimate the gauge factor in the highly doped 3C-SiC. For strain-free n-type 3C-SiC, the the number of electrons in each equivalent valley is given by: 
\begin{equation}
n_0= N_{c}\times \mathcal{F}_{1/2}(\frac{E_c-E_F}{k_BT})
\end{equation} 
\noindent where $N_{c}$ is the effective density of state in the conductance band; $\mathcal{F}_{1/2}$ is the Fermi-Diract integral; $E_c$ and $E_F$ are the energy of each valley in the conduction band and the Fermi level; $k_B$ is the Boltzmann constant; and $T$ is the absolute temperature. Under applied strain $\varepsilon$ in [100] orientation, the energy of the valley located in [100] axis will shift up while the energies in the other valleys in [010] and [001] axes will shift down an amount of $\Delta E = \Xi_u \varepsilon$. Here, $\Xi_u$ is an independent constant of the deformation energy. As a result, the change in electron concentration in each valley is:
\begin{equation}
\begin{aligned}
\Delta n &= N_c\times \{\mathcal{F}_{1/2}(\frac{E_c \pm \Delta E -E_F}{k_BT})-\mathcal{F}_{1/2}(\frac{E_c- E_F}{k_BT})\}\\
&\approx N_c\times \frac{\pm\Delta E}{k_BT} \times \mathcal{F}_{-1/2}(\frac{E_c-E_F}{k_BT}) 
\end{aligned}
\end{equation}
where $\mathcal{F}_{-1/2} = \delta\mathcal{F}_{1/2}/\delta E$. Consequently, based on the electron transfer phenomenon, the gauge factor of n-type 3C-SiC is given by \cite{Kanda2,Toriyama,Phan_review}:
\begin{equation}
G= 1+2\nu -\frac{\Xi_u(L-1)}{3k_{B}T(2L+1)}\times (2+\nu')\times \frac{\mathcal{F}_{-1/2}}{\mathcal{F}_{1/2}}
\label{eq:GF}
\end{equation}
\noindent where $L=\mu_\perp /\mu_\parallel$ is the ratio of the transverse and longitudinal electron mobilities, and $\nu$ and $\nu '$ are the Poisson ratio of 3C-SiC and the substrate, respectively. The Fermi-Dirac integral was calculated based on Chang-Elizabeth approximation \cite{Chang}, while the  deformation energy $\Xi_u$ = 6.3eV was obtained based on the theoretical work of Lambrecht \emph{et al.} on full potential band calculation \cite{Lambrecht}. Additionally, the mobility ratio L = 2.74 was obtained from the experimental work reported in \cite{Kaplan}. Substituting these values into Eq. \ref{eq:GF}, the gauge factor in degeneracy doped 3C-SiC with carrier concentration of $5\times 10^{19}$ cm$^{-3}$ to 10$^{20}$ cm$^{-3}$ varies from -8 to -14 which is consistent with our experimentally measured value. This result indicates that the electron transfer mechanism can qualitatively and quantitatively explain the piezoresistive effect in the highly doped 3C-SiC. 

 In conclusion, this work reports the piezoresistive effect in highly doped 3C-SiC nanothin film transferred onto a 6-inch glass substrate, employing anodic bonding technique. Experimental data shows a negative gauge factor of approximately -8.6 in the longitudinal [100] direction of the n-type 3C-SiC, which is in solid agreement with numerical analysis based on electron transfer effect in cubic materials under stress. The large gauge factors, superior mechanical properties, excellent chemical inertness, and bio-compatibility of the transferred 3C-SiC on glass could open up promising opportunities for MEMS applications in harsh environments, as well as bio-sensing. 
 

 This work was partially funded by the linkage grants LP150100153 and LP160101553 from the Australian Research Council (ARC). This work was performed in part at the Queensland node of the Australian National Fabrication Facility, a company established under the National Collaborative Research Infrastructure Strategy to provide nano and micro-fabrication facilities for Australia's researchers.  H.P. Phan acknowledges research grants from the Australian Nanotechnology Network Overseas Travel Fellowship and Griffith University Postdoctoral Fellowship (GUPF).

\end{document}